\documentclass[twocolumn,showpacs,prb]{revtex4}
\usepackage{graphicx}
\usepackage{amsmath}
\usepackage{bm}
\begin{document}
\title{Brownian motion and diffusion: from stochastic processes to
chaos and beyond}
\author{F.~Cecconi$^1$, M.~Cencini$^1$, M.~Falcioni$^1$ and A.~Vulpiani$^{1,2}$} 
\affiliation{$^1$Center for Statistical Mechanics and Complexity, INFM Roma-1 \\ 
Dip.~di~Fisica, Universit\`a di Roma ``La Sapienza''\\
P.le Aldo Moro, 2 I-00185 Roma, Italy \\
$^2$INFN Sezione di Roma ``La Sapienza''}
\date{\today}

\begin{abstract}
One century after Einstein's work, Brownian Motion still remains both a
fundamental open issue and a continous source of inspiration 
for many areas of natural sciences. 
We first present a discussion about stochastic and deterministic approaches
proposed in the literature to model the Brownian Motion
and more general diffusive behaviours.  
Then, we focus on the problems concerning   
the determination of the microscopic nature of diffusion by means of 
data analysis. Finally, we discuss the general conditions required
for the onset of large scale diffusive motion.
\end{abstract}


\maketitle

{\bf Brownian motion (BM) played a fundamental role in the development
of molecular theory of matter, statistical mechanics and stochastic
processes.  Remarkably, one century after Einstein's work, 
BM is still at the origin of scientific discussions as testified by a recent 
experiment performed to detect a trace of deterministic chaotic sources on 
macroscopic diffusion.    
Several authors, which discussed the results of such an experiment, 
argued that the possibility to discern experimentally between 
a deterministic chaotic and noisy dynamics, 
at the microscopic level, is severely limited by subtle technical and 
conceptual points. However, the remarks raised by the scientific community 
have gone over the criticism and have led to a deeper understanding of the
role of chaos in the diffusion.
 
After a short historical introduction to BM, we
focus on the dynamical conditions to observe macroscopic
diffusion. In particular, we discuss the technical and conceptual
limits in distinguishing, by means of data analysis, the deterministic 
or stochastic nature of diffusion. A main tool for that is the
$\epsilon$-entropy.  Part of the discussion is devoted to the problem
of macroscopic diffusion in deterministic non-chaotic dynamics.}

\section{Introduction}
\label{sec:intro}
At the beginning of the twentieth century, the atomistic theory of
matter was not yet fully accepted by the scientific community. 
While searching for phenomena that would prove, beyond any doubt, 
the existence of atoms, Einstein realized that ``{\it ... according
to the molecular-kinetic theory of heat, bodies of
microscopically-visible size suspended in a liquid will perform
movements of such magnitude that they can be easily observed in a
microscope ...}'', as he wrote in his celebrated paper in 1905 
~\cite{e1905}.  
In this work, devoted to explain the irregular motion of Browinan particles 
on theoretical grounds,  Einstein argued that the motion of these small 
bodies has a diffusive character. Moreover, he discovered an important relation
involving the diffusion coefficient $D$, the fluid viscosity $\eta$,
the particles radius $a$ (having assumed spherical particles),
Avogadro's number $N_A$, the temperature $T$ and the gas constant $R$:
\begin{equation}
D = \frac{1}{N_A}  \frac{RT}{6\pi \eta a} \,\, .
\label{diff}
\end{equation}
This relation can be employed, and actually had been,
to determine experimentally the Avogadro's 
number~\cite{chandra}.
Indeed, the diffusion coefficient can be measured by monitoring the 
growth, with time $t$, of the particle displacement 
$\Delta x=x(t)-x(0)$, which is expected to behave as  
\begin{equation}
\langle (\Delta x)^2 \rangle \simeq 2\, D \, t \, \,.
\end{equation}
Einstein relation (\ref{diff}), that may be
seen as the first example of the fluctuation-dissipation
theorem~\cite{kubo}, allowed for the determination of Avogadro's
number and gave one of the ultimate evidences of the existence of atoms. 
 
Einstein's theoretical explanation of BM is based on the intuition that the 
irregular motion of a Brownian particle is a consequence of the huge 
number of collisions per unit time with the surrounding fluid
molecules. Since Einstein's approach, diffusion and 
irregular phenomena were commonly associated to the 
presence of many degrees of freedom. 
The effects of the disregarded degrees of
freedom on an observed small part of a system can be either studied
directly (as initiated by Smoluchowski~\cite{Smoluch}) or modeled by
means of stochastic dynamics (as proposed by Langevin~\cite{lange}).
From the latter point of view, BM provided the first and main stimulus
to the building of the modern theory of stochastic processes.
 
After the (re)discovery of deterministic chaos \cite{Lorenz,Ottbook},
it was clear that also fully deterministic and low dimensional systems
can give rise to erratic seemingly random motions, practically
indistinguishable from those produced by a stochastic process. 
This implied an affective unpredictability of chaotic systems and the
need for a probabilistic description also of a strictly deterministic
world.
The success in understanding the basic mechanisms for the onset of
chaos, and the wealth of interesting phenomena occurring in low
dimensional systems hinted at optimistic expectations about the
possibility of a systematic deterministic approach to irregular
natural phenomena.  This rised a rapid development of time
series analysis with the idea to demonstrate the deterministic
character of many irregular phenomena.

Nowadays we are aware of the limits of this optimistic
programme~\cite{Abarnabel,Kantzbook}, and we know that a 
definite answer on the deterministic or stochastic character of
experimental signals is impossible. However, some tools of time 
series analysis, such as the entropy analysis at varying the scale of
resolution, are very useful to characterize important features 
of complex systems.  
Among the recent developments in this context, we can
mention the experiment by Gaspard et al.~\cite{gaspard} on the motion
of a Brownian particle. The debate~\cite{dettman, grass} around the
possible theoretical interpretation of the experiment is a clear
indication of how, one century after the seminal Einstein's work, 
BM continues to be a subject of intricate and fascinating discussions.
 
Beyond its undoubted importance for applications in many natural
phenomena, deterministic chaos also enforces us to reconsider some basic
problems standing at the foundations of statistical mechanics such as, 
for instance, the applicability of a statistical description to
low dimensional systems. 

In addition, the combined effects of noise and deterministic evolution can 
generate highly nontrivial and rather intriguing behaviours. As an
example, we just mention the stochastic resonance \cite{Benzi,Gamma} and
the role of colored noise in dynamical systems.\cite{Hanngi} 
 
The aim of this paper is a discussion on the viable approaches to
characterize and understand the dynamical (microscopic) character of
BM (Sect.~II). In particular, we shall focus on the distinction
between chaos and noise from a data analysis and on conceptual aspects
of the modeling problem (Sect.~III). Moreover, we shall investigate
and discuss about the basic microscopic ingredients necessary for BM
as, for instance, the possibility of genuine BM in non-chaotic
deterministic systems (Sect.~IV). Finally, we conclude (Sect.~V) with
a discussion on the role of chaos in statistical mechanics.

\section{The origin of diffusion}
\label{sec:diff}
Einstein's work on BM is based on 
statistical mechanics and thermodynamical
considerations applied to suspended
particles, with the assumption of velocity decorrelation
(molecular chaos).
  
Instead, one of the first attempts to develop a purely dynamical theory of BM 
dates back to Langevin~\cite{lange} that, as himself writes, gave ``{\it ... a
demonstration} [of Einstein results] {\it that is infinitely more
simple by means of a method that is entirely different.}''  Langevin
considers the Newton equation for a small spherical particle in a fluid, taking
into account that the Stokes viscous force it experiences is only a
mean force. In one direction, say e.g. the $x$-direction, one has:
\begin{equation}
m \,\, \frac{d^2 x}{dt^2} = -6 \pi \eta a \, \, 
\frac{d x}{dt} + F
\label{langevin}
\end{equation} 
where $m$ is the mass of the particle.  In the r.h.s. the first term
is the Stokes viscous force.  $F$ is a fluctuating random force
which models the effects of the huge number of impacts with the
surrounding fluid molecules, responsible for the thermal
agitation of the particle. In statistical mechanics terms,
this corresponds to molecular chaos.  
 
With the assumption that the force $F$ is a Gaussian, time
uncorrelated random variable, the probability distribution functions
(pdf) for the position and velocity of the Brownian particle can be
exactly derived~\cite{uhle}.  In particular, the pdf of the position,
at long times, reduces to the Gaussian distribution in agreement with
Einstein's result.
 
Langevin's work along with that of Ornstein and Uhlenbeck~\cite{uhle}
are at the foundation of the theory of stochastic differential
equations. The stochastic approach is however unsatisfactory
being at the level of a phenomenological description.
 
The next theoretical challenge toward the building of a dynamical
theory of Brownian motion is to understand its microscopic origin from
first principles.  A very early attempt was made in 1906 by
Smoluchowski, who tried to derive the large scale diffusion of
Brownian particles starting from the microscopic description of their
collisions with the fluid molecules~\cite{Smoluch}.  A renewed
interest on the subject appeared some years later, when it was
realized that even purely deterministic systems composed of a large
number of particles produce macroscopic diffusion, at least on finite
time scales.  These models had an important impact in the
justification of Brownian motion theory and, more in general, in
deriving a consistent microscopic theory of irreversibility.
 
Some of these works considered chains of harmonic oscillators of equal
masses~\cite{turner1,mazur,ford1,phill}, while 
others~\cite{turner2,rubin,mazur2} analyzed the motion of a heavy
impurity linearly coupled to a chain of equal mass oscillators.  For
an infinite number of oscillators, the momentum of the heavy
particle behaves as a genuine stochastic process described by the
Langevin equation~(\ref{langevin}).  When their number is finite,
diffusion remains an effective phenomenon lasting for a (long but)
limited time.
 
Soon after the discovery of dynamical chaos, it was realized that also
simple low dimensional deterministic systems may exhibit a diffusive
behavior.  In this framework, the two-dimensional Lorentz
gas~\cite{lorentz}, describing the motion of a free particle through a
lattice of hard round obstacles, provided the most valuable example.
Particle trajectories can be ballistic (with very few collisions in
the case of infinite horizon) or chaotic as a consequence of the
convexity of the obstacles. In the latter case, at large times, the
mean square displacement from the particle initial condition grows
linearly with time. Lorentz system
is closely related to the Sinai billiard~\cite{sinai1,sinai2}, which
can be obtained from the Lorentz gas by folding the trajectories into
the unitary lattice cell. The extensive study on billiards
has shown that chaotic behavior might usually be associated to diffusion
in simple low dimensional models, supporting the idea that chaos
was at the very origin of diffusion.  However, more recently (see
e.g. Ref.~\onlinecite{DC}) it has been shown that even non-chaotic
deterministic systems, such as a bouncing particle in a
two-dimensional billiard with polygonal but randomly distributed
obstacles (wind-tree Ehrenfest model), may exhibit a diffusion-like
behavior (see Sect.~IV).
 
Deterministic diffusion is a generic phenomenon present
also in simple chaotic maps on the line.  Among the many contributions
we mention the work by Fujisaka, Grossmann~\cite{fuji,gross} and
Geisel~\cite{gei1,gei2}.  A typical example is the
$1d$ discrete-time dynamical system:
\begin{equation}
x(t+1) = [x(t)] + F(x(t)-[x(t)]) \,  ,
\label{eq:chaos}
\end{equation}
where $x(t)$ (the position of a point-like particle) performs diffusion
in the real axis.  The bracket $[\dots]$ denotes the integer part of
the argument. $F(u)$ is a map defined on the interval $[0,1]$ that
fulfills the following properties:\\
\indent i) The map, $u(t+1)=F(u(t))$ (mod $1$) is chaotic.\\
\indent ii) $F(u)$ must be larger than $1$ and smaller than $0$ for
some values of $u$, so there exists a non vanishing probability to escape
from each unit cell (a unit cell of real axis is
every interval $C_{\ell} \equiv [\ell,\ell+1]$, with $\ell \in {\bf Z}$).\\
\indent iii) $F_r(u)=1-F_l(1-u)$, where $F_l$ and $F_r$ define the map
in $u\in [0,1/2[$ and $u\in [1/2,1]$ respectively. This anti-symmetry
condition with respect to $u= 1/2$ is introduced to avoid a net drift.
 
A very simple and much studied example of $F$ is 
\begin{eqnarray}
F(u) =
\left\{
\begin{array}{ll}
2(1+a) u   \qquad \qquad \quad \, \, \mbox{if~}  u \in [0,1/2[   \\
2(1+a) (u-1) + 1 \,\quad\mbox{if~}  u \in [1/2,1]
\end{array}\right.
\label{eq:chaos2}
\end{eqnarray}
where $a>0$ is the control parameter.
It is useful to remind the link between diffusion and velocity
correlation, i.e. the Taylor-Kubo formula, that helps in understanding
how diffusion can be realized in different ways.
Defining $C(\tau)=\langle v(\tau) v(0) \rangle$ as the 
velocity correlation function, where $v(t)$ is the velocity of the 
particle at time $t$. 
It is easy to see that for continuous time systems (e.g. Eq.(\ref{langevin}))
\begin{equation}
\langle \left( x(t)-x(0)\right)^2 \rangle 
\simeq  2\; t\; \int_0^t 
d\tau \; C(\tau)\,.
\label{pippo1}
\end{equation}
Standard diffusion, with  $D= \int_0^{\infty} \; d \tau \;
C(\tau)$, is always obtained whenever the hypotheses for the
validity of the central limit theorem are verified: \\
\indent I) the variance of the velocity must be finite: 
$\langle v^2 \rangle < \infty$

\indent II) the decay to zero of the velocity correlation 
function $C(\tau)$ at large times  should be faster than $ \tau^{-1}$.

In discrete-time systems, the velocity $v(t)$ and the integration of 
$C(t)$ are replaced by the finite difference $x(t+1) - x(t)$ and   
by the quantity $\langle v(0)^2 \rangle/2 + \sum_{\tau} C(\tau)$ 
respectively. 
 
Condition I) is justified by the fact that having an infinite
variance for the velocity is rather unphysical. It should be noted that
this requirement is independent of the microscopic dynamics under
consideration: Langevin, deterministic chaotic or regular dynamics. 
 
Condition II), corresponding to the request of molecular chaos, is
surely verified for the Langevin dynamics where the presence of the
stochastic force entails a rapid decay of $C(\tau)$. In deterministic
regular systems, such as the many oscillator model, the velocity
decorrelation comes from the huge number of degrees of freedom that
act as a heat bath on a single oscillator.  While in the (non-chaotic)
Ehrenfest wind-tree model decorrelation originates from the disorder
in the obstacle positions, the situation is more subtle for
deterministic chaotic systems. In fact, even if nonlinear
instabilities generically lead to a memory loss and, henceforth, to
the validity of the molecular-chaos hypothesis, slow decay of
correlation , e.g. $ C(\tau) \sim \tau^{-\beta}$ with $\beta < 1$, may
appear in very intermittent systems~\cite{R1}.  When this happens,
condition II) is violated, and superdiffusion, $\langle x^2(t) \rangle
\sim t^{2-\beta}$, is observed.  Though interesting, superdiffusion is
a quite rare phenomenon. Moreover, usually, small changes of the
control parameters of the dynamics restore standard diffusion.
Therefore, also for chaotic systems we can state that the ``rule'' is
the standard diffusion and the ``exception'' is the
superdiffusion~\cite{R2}.

We end this section by asking whether is it possible to determine, by
the analysis of a Brownian particle, if the microscopic dynamics
underlying the observed macroscopic diffusion is stochastic, deterministic
chaotic or regular?

\section{Distinction between Chaos and Noise}
Inferring the microscopic deterministic character of Brownian motion
on an experimental basis would be attractive from a fundamental
viewpoint. Moreover it could provide further evidence to some recent
theoretical and numerical studies~\cite{PH88,BDPD97}.  Before
discussing a recent experiment\cite{gaspard} in this direction, we
must open the ``Pandora box'' of the longstanding and controversial
problem of distinguishing chaos from noise in signal
analysis~\cite{cenc}.
 
The first observation is that, very often in the analysis of experimental 
time series,  there is not a unique model of the
``system'' that produced the data. Moreover, even the knowledge of the
``true'' model might not be an adequate answer about the character of
the signal. From this point of view, BM is a paradigmatic example: in
fact it can be modeled by a stochastic as well as by a deterministic
chaotic or regular process.
 
In principle a definite answer exist. If we were able to determine the
maximum Lyapunov exponent ($\lambda$) or the Kolmogorov-Sinai entropy
($h_{KS}$) of a data sequence, we would know without uncertainty
whether the sequence was generated by a deterministic law ($\lambda ,
h_{KS} < \infty$) or by a stochastic one ($\lambda ,h_{KS} \to
\infty$). Nevertheless, there are unavoidable practical limitations in
computing such quantities. Those are indeed defined as infinite time
averages taken in the limit of arbitrary fine resolution. But, in
experiments, we have access only to a finite, and
often very limited, range of scales and times.
 
However, there are measurable quantities that are appropriate for
extracting information on the signal character. In particular, we
shall consider the $(\epsilon,\tau)$-entropy per unit
time~\cite{kolmogorov,shannon,Gaspard93c} $h(\epsilon,\tau)$ that
generalizes the Kolmogorov-Sinai entropy. In a nutshell, while for
evaluating $h_{KS}$ one has to detect the properties of a system with
infinite resolution, for $h(\epsilon, \tau)$ a finite scale
(resolution) $\epsilon$ is requested. The Kolmogorov-Sinai entropy is
recovered in the limit $\epsilon \to 0$, i.e. $h(\epsilon,\tau) \to
h_{KS}$. This means that if we had access to arbitrarily small scales,
we could answer the original question about the character of the law
that generated the recorded signal. Even if this limit is
unattainable, still the behavior of $h(\epsilon,\tau)$ provides a
very useful scale-dependent description of the nature of a signal.

\subsection{$\epsilon$-entropy}
The $\epsilon$-entropy was originally introduced in the context of
information theory by Shannon~\cite{shannon} and, later, by
Kolmogorov~\cite{kolmogorov} in the theory of stochastic processes.
An operative definition is as follows.
 
One considers a continuous variable ${\bf x}(t) \in \Re^d$, that
represents the state of a $d$-dimensional system, and one introduces
the vector
\begin{equation}
\label{eq:2-1}
{\bf X}^{(m)}(t)= \left( {\bf x}(t), \dots, 
{\bf x}(t+m\tau-\tau) \right)\,, 
\end{equation}
which lives in $\Re^{md}$ and 
is a portion of the trajectory discretized in time with step $\tau$. 
Then the phase space $\Re^d$ is
partitioned using hyper-cubic cells of side $\epsilon$. The vector 
${\bf X}^{(m)}(t)$ is coded into the word, of length $m$,
\begin{equation}
\label{eq:2-2}
W^{m}(\epsilon, t) = \left( i(\epsilon, t), 
\dots, i(\epsilon, t+m \tau -\tau) \right)\,,
\end{equation}
where $i(\epsilon, t+j \tau)$ labels the cell in $\Re^d$ containing
${\bf x}(t+j \tau)$. For bounded motions, the number of visited cells
(i.e.  the alphabet) is finite. Under the hypothesis of
stationarity, the probabilities $P(W^{m}(\epsilon))$ of the admissible 
words $\lbrace W^{m}(\epsilon) \rbrace$ are obtained 
from the time evolution of ${\bf X}^{(m)}(t)$.  
The
$(\epsilon,\tau)$-entropy per unit time, $h(\epsilon , \tau)$ is
then defined by~\cite{shannon}:
\begin{equation}
\label{eq:2-3b}
h(\epsilon , \tau)=\lim _{m \to \infty} h_m(\epsilon , \tau) 
= {1 \over \tau} 
\lim _{m \to \infty} {1 \over m} H_{m} (\epsilon,\tau) ,
\end{equation} 
where $H_m$ is the $m$-block entropy:
\begin{equation}
\label{eq:2-4}
H_{m} (\epsilon,\tau) = - \sum _{ \lbrace W^{m}(\epsilon) \rbrace } 
P(W^{m}(\epsilon))  \ln P(W^{m}(\epsilon))\,,
\end{equation} 
and $h_m(\epsilon , \tau)= \lbrack H_{m+1} (\epsilon,\tau) -H_m
(\epsilon,\tau) \rbrack/\tau$. 
\\\indent
It is worth remarking a few important points.  A rigorous mathematical
procedure\cite{Gaspard93c} would require to take the infimum over all 
possible partitions 
with elements of size smaller than $\epsilon$.
The Kolmogorov-Sinai entropy is obtained in the limit of small $\epsilon$
\begin{equation}
\label{eq:2-5}
h_{KS} = \lim_{\epsilon \to 0} h(\epsilon, \tau)\;.
\end{equation} 
Note that in deterministic systems $h(\epsilon)$ and henceforth
$h_{KS}$ do not depend on the sampling time so that (\ref{eq:2-5}) can
be in principle used with any choice of $\tau$~\cite{ER}. However, in
practical computations, where the rigorous definition is not
applicable, the specific value of $\tau$ is important and
$h(\epsilon)$ may also depend on the used norm.  For very small
$\epsilon$, no matter of the norm, the correct value of the
Kolmogorov-Sinai entropy is usually recovered. Indeed, when the
partition is very fine, usually, it well approximates a generating
partition.  It is worth reminding that the Kolmogorov-Sinai entropy is
a dynamical invariant, i.e. independent of the used state
representation (\ref{eq:2-1}).

In deterministic systems the following chain of inequalities holds~\cite{ER}
\begin{equation}
h(\epsilon,\tau) \leq h_{KS} \leq \sum_i^{+}\lambda_i\,,
\label{eq:ineq}
\end{equation}
where the summation is over all positive Lyapunov exponents. 
A system is chaotic when $0< h_{KS} < \infty$ and regular 
when $h_{KS}=0$. Typically, one observes that $h(\epsilon,\tau)$ 
attains a plateau $h_{KS}$, below a resolution threshold, $\epsilon_c$, 
associated to the smallest characteristic length scale of the system. 
Instead, for $\epsilon>\epsilon_c$ due to (\ref{eq:ineq})
$h(\epsilon,\tau)<h_{KS}$, in this range the details of the
$\epsilon$-dependence may be informative on the large scale (slow)
dynamics of the system (see e.g. Refs.~\onlinecite{cenc,rass}).

In stochastic signals $h_{KS}=\infty$, but for any $\epsilon>0$,
$h(\epsilon,\tau)$ is finite and a well defined function of $\epsilon$ and 
$\tau$. The nature of the 
dependence of $h(\epsilon,\tau)$ on $\epsilon$ and $\tau$ provide a
characterization of the underlying stochastic process (see
Refs.~\onlinecite{kolmogorov,Gaspard93c,rass}).  For an important and
wide class of stochastic processes~\cite{Gaspard93c} an
explicit expression  for $h(\epsilon, \tau)$ can be given in the limit
$\tau \to 0$. This is the case of stationary Gaussian processes
characterized by a power spectrum $S(\omega)\propto
\omega^{-(2\alpha+1)}$, with $0 < \alpha <1$, for
which~\cite{kolmogorov}:
\begin{equation}
\lim_{\tau\to 0} h(\epsilon,\tau) \sim {\epsilon^{-1/\alpha}} \; .
\label{eq:kolmo56}
\end{equation}
The case $\alpha=1/2$, corresponding to the power spectrum of a
Brownian signal, would give $h(\epsilon)\sim \epsilon^{-2}$. Some
stochastic processes, such as {\em e.g.} time uncorrelated and bounded
ones, are characterized by a logarithmic divergence below a critical
$\epsilon_c$, which may depend on $\tau$.

\subsection{Numerical determination of the $\epsilon$-entropy}
In experiments, usually, only a scalar variable $u(t)$ can be measured
and moreover the dimensionality of the phase space is not known. 
In these cases it is
reconstructed by delay embedding 
technique~\cite{Kantzbook,Abarnabel}, where the 
vector ${\bf X}^{(m)}(t)$ is
build as $\left( u(t), u(t+\tau), \dots, u(t+m\tau-\tau) \right)$, now
in $\Re^m$. This is a special instance of (\ref{eq:2-1}).
\begin{figure}
\includegraphics[draft=false, scale=0.67, clip=true]{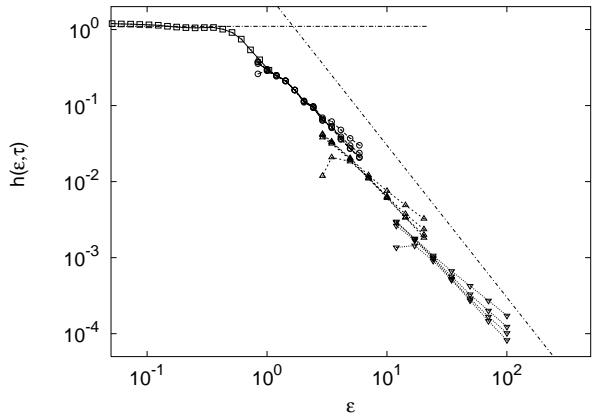}
\caption{ Numerically evaluated $(\epsilon,\tau)$-entropy for the map
(\ref{eq:mappa}) with $p=0.8$ computed by the standard techniques
$[6]$ at $\tau=1$ ($\circ$), $\tau=10$ ($\bigtriangleup$) and
$\tau=100$ ( $\bigtriangledown$) and different block length
($m=4,8,12,20$).  The boxes give the entropy computed with $\tau=1$ by
using periodic boundary condition over $40$ cells.  The straight lines
correspond to the two asymptotic behaviors, $h(\epsilon)=h_{KS}\simeq 1.15$
and $h(\epsilon) \sim \epsilon^{-2}$.}
\label{fig:1}
\end{figure}

Then to determine the entropies $H_m(\epsilon)$, very efficient
numerical methods are available~\cite{CP,GP} (the reader may find an
exhaustive review in Refs.~\onlinecite{Kantzbook,Abarnabel}). Here,
avoiding technicalities, we just mention some subtle points which
should be taken into account in data analysis.
\\\indent
First, if the information dimension of the attractor for a given
system is $d_1$ then, to have a meaningful measure of the entropy, the
embedding dimension $m$ has to be larger than $d_1$.  Second, as above
mentioned, the plateau $h_m(\epsilon)\approx h_{KS}$ appears only
below a critical $\epsilon_c$, meaning that it is possible to
distinguish a deterministic signal from a random one only for
$\epsilon<\epsilon_c$. However, one should be aware of the fact that
the finiteness of the data set imposes a lower cut-off scale
$\epsilon_d$ below which no information can be extracted from the data
(see Ref.~\onlinecite{Olbrich97}).  Also in stochastic signals, there
exists a lower critical cut-off $\epsilon_d$ due to the finiteness of
the data set, and often, as mentioned at the end of the previous
subsection, one has logarithmic divergences below $\epsilon_c$.
Since, this is also what happens in general for $\epsilon<
\epsilon_d$, the interpretation of the results requires much
attention~\cite{Olbrich97,Kantzbook,cenc}.
\\\indent
Therefore, if $m$ is not large enough and/or $\epsilon$ is not small
enough (or in the lack of a good estimation of the important range of
scales for the different behaviors) one may obtain misleading results.
\\\indent
Another important problem concerns the choice of $\tau$. If $\tau$ is
much larger or much shorter than the characteristic time-scale of the
system at the scale $\epsilon$, then the correct behavior of the
$\epsilon$-entropy~\cite{cenc} cannot be properly recovered.
\\\indent
To exemplify the above difficulties let us consider the map
\begin{equation}
x(t+1)=x(t) + p\sin(2\pi x(t))\, ,
\label{eq:mappa}
\end{equation}
which, for $p > 0.7326\dots$, is  chaotic  (similarly to (\ref{eq:chaos}))
and displays large scale diffusion. On the basis of the previous discussion, 
the $\epsilon$-entropy is expected to behave as
\begin{equation}
\label{eq:3-2}
h(\epsilon) \simeq 
\left\{ 
\begin{array}{ll}
\lambda \,\,\,{\rm for} \,\,\, \epsilon \ll 1 \\
                                                     \\
D / \epsilon ^2 \,\,\, {\rm for} \,\,\, \epsilon \gg 1  
\end{array}
\right.\;, 
\end{equation} 
where $\lambda$ is the Lyapunov exponent and $D$ is the diffusion
coefficient. The typical problems encountered in numerically computing
$h(\epsilon)$ can be appreciated in Fig.~\ref{fig:1}.  First
notice that the threshold $\epsilon_c \approx 1$.  
As for the importance of the
choice of $\tau$, note that the diffusive behavior $h(\epsilon)\sim
\epsilon^{-2}$ is roughly obtained only by considering the envelope of
$h_m(\epsilon,\tau)$ evaluated at different values of $\tau$. The reason 
for this is as follows.
The natural sampling interval would be $\tau =1$, however this choice 
requires considering larger and larger embedding dimensions $m$ at 
increasing $\epsilon$. 
Indeed, a simple dimensional argument suggests that 
the characteristic time of the system is determined by its diffusive 
behaviour $T_\epsilon \approx \epsilon^2 / D$.
If we consider for example,
$\epsilon = 10$ and the typical values of the diffusion coefficient $D
\simeq 10^{-1}$, the characteristic time, $T_{\epsilon}$, is much larger
than the elementary sampling time $\tau=1$. On the other hand, the
plateau at the value $h_{KS}$ can be recovered only for $\tau\approx
1$, even if, in principle, any value of $\tau$ could be used.
\\\indent
The above difficulties can be partially overcome by means of a recently
introduced method based on exit times~\cite{PhysicaDnostro}.  The
main advantage of this approach is that it is not needed to fix {\it a
priori} $\tau$, because the ``correct'' $\tau$ is automatically
selected.

\subsection{Does Brownian motion arise from chaos, noise or regular 
dynamics?}
We are now ready to discuss the experiment and its results 
reported in Ref.~\onlinecite{gaspard}. In this experiment, 
a long time record (about $1.5 \times
10^5$ data points) of the motion of a small colloidal particle in
water was sampled at regular time intervals ($\Delta t=1/60$ s) with a
remarkable high spatial resolution ($25$ nm).  To our knowledge, this
is the most accurate measurement of a BM. The data were then processed
by means of standard nonlinear time-series analysis tools, i.e. the
Cohen-Procaccia method~\cite{CP}, to compute the $\epsilon$-entropy.
This computation shows a power-law dependence $h(\epsilon)\sim
\epsilon^{-2}$. Actually, similarly to what displayed in
Fig.~\ref{fig:1}, this behavior is recovered only by considering the
envelope of the $h(\epsilon,\tau)$-curves, for different $\tau$'s. 
However, unlike to Fig.~\ref{fig:1}, no  
saturation $h(\epsilon,\tau)\approx \mbox{const}$ is observed in the 
small $\epsilon$-region because of the finiteness of data set and 
resolution as well. 
From the previous discussion, this can be understood as   
the fact that $\epsilon_c$ of the observed system is much smaller than  
the smallest detectable scale extracted from data and therefore the 
$KS$-entropy cannot be properly estimated. 
Nevertheless, from the chain of inequalities (\ref{eq:ineq}) and 
by {\it assuming} from the outset that the system dynamics is 
deterministic, the authors deduce, from the positivity
of $h(\epsilon)$, the existence of positive Lyapunov exponents in the
system.  Their conclusion is thus that microscopic chaos is at the
origin of the macroscopic diffusive behavior.
\\\indent
As pointed out by several works, a few  points
need to be considered in the data analysis of the aforementioned
experiment, namely: the huge amount of involved degrees of freedom
(Brownian particle and the fluid molecules); the impossibility to
reach high enough (spatial and temporal) resolution; the limited
amount of data points.
\\\indent
The limitation induced by the finite resolution is particularly
relevant to the experiment. Even if one assumes that the number of
data points and the embedding dimension are large enough, the
impossibility to see a saturation to a constant value,
$h(\epsilon) \approx \mbox{const}$, prevents any conclusion 
about the character of the analyzed signal.  For example, whenever the 
the analysis of Fig.~\ref{fig:1} would be restricted to the region 
with $\epsilon>1$, then discerning whether the data were originated by
a chaotic system or by a stochastic process would be impossible. 
In fact in both cases the
behavior $h(\epsilon)\sim \epsilon^{-2}$ would have been observed.
\begin{figure}[t!]
\includegraphics[draft=false, scale=0.65, clip=true]{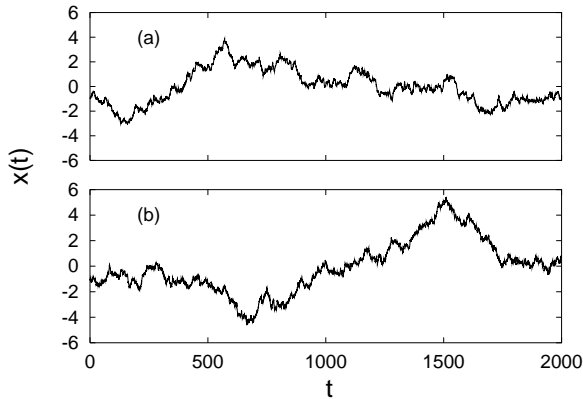}
\caption{(a) Signals obtained from Eq.~(\ref{eq:mazur}) with $M=10^4$
and random phases uniformly distributed in $[0,2\pi]$. The numerically
computed diffusion constant is $D \approx 0.007$.  (b) Time record
obtained with a continuous random walk (\ref{eq:RW}) with the same
value of the diffusion constant as in (a).  In both cases data are
sampled with $\tau=0.02$, i.e. $10^5$ data points.}
\label{fig:2}
\end{figure}
%
\\\indent
As for the number of degrees of freedom, we recall that, for a correct
evaluation of the entropy of the microscopic dynamics, a very high
embedding dimension should be used, in practice $m>d_1$.  For a fluid
(with $O(10^{23})$ molecules) the necessary number of points
is of course prohibitive.  Moreover, as stressed by Grassberger and
Schreiber~\cite{grass}, when the number of degrees of freedom is so
high that it can be considered practically infinite there is an
additional difficulty related to the definition of entropy and
Lyapunov exponents, which become norm dependent.
\\\indent
Furthermore, the limited amount of data severely affects even our
ability to recognize if the signal is deterministic but of zero
entropy (i.e. regular).  This has been pointed out by Dettmann et
al.~\cite{dettman,DC}, who have shown that the same entropic analysis
of Ref.~\onlinecite{gaspard}, applied to the Ehrenfest wind-tree model 
(see next section) reproduced results
very similar to those extracted in the Brownian experiment. This
model is deterministic and non-chaotic. In fact if the time records
were long enough to see the periodic nature of the signal, and the
embedding dimension high enough to resolve the system manifold, the
measured entropy would have been zero.
\\\indent
%
\begin{figure}[t!]
\includegraphics[draft=false, scale=1, clip=true]{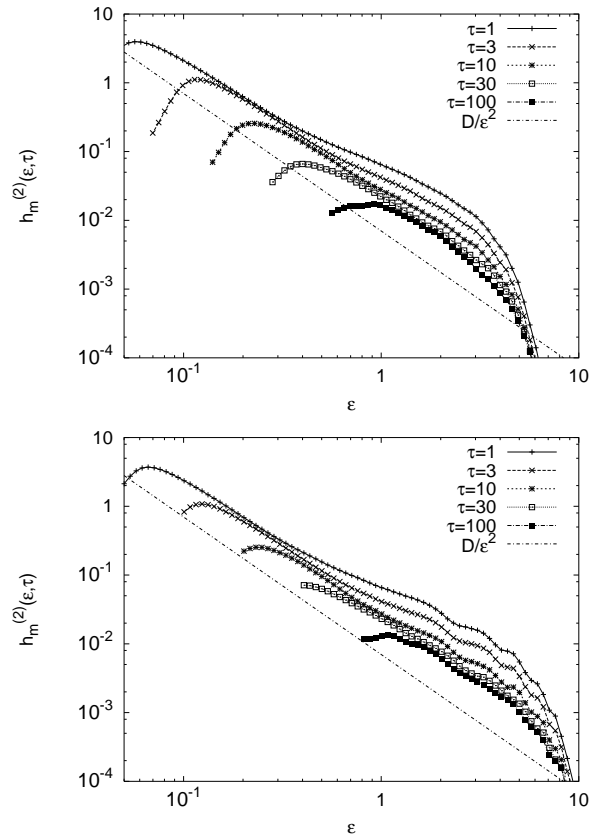}
\caption{$h(\epsilon)$ computed with the Grassberger-Procaccia
algorithm using using $10^5$ points from the time series of
Fig.~\ref{fig:2}.  We show the results for embedding dimension
$m=50$.  The straight-lines show the $D/\epsilon^2$ behavior.}
\label{fig:3}
\end{figure}
The following example serves as a clue to better understand these
points.  Let us consider two signals, the first generated by 
a continuous random walk:
\begin{equation}
\dot{x}(t)=\sqrt{2D} \eta(t)\,,
\label{eq:RW}
\end{equation}
where $\eta$ is a zero mean Gaussian variable with $\langle
\eta(t)\eta(t') \rangle = \delta(t-t')$, and the second obtained as
a superpositions of Fourier modes: 
\begin{equation}
x(t)= \sum_{i=1}^M X_{0i}\sin \left(\Omega_i t+\phi_i \right) \,.
\label{eq:mazur}
\end{equation}
The coordinate $x(t)$ in Eq.~(\ref{eq:mazur}), 
upon properly choosing the frequencies~\cite{mazur2,cenc} and the amplitudes (e.g. $X_{0i}\propto \Omega_i^{-1}$), describes the motion 
of a heavy impurity in a chain of $M$ linearly coupled harmonic 
oscillators.  
We
know~\cite{mazur2} that $x(t)$ performs a genuine BM in the limit
$M\to \infty$. For $M<\infty$ the motion is periodic and regular, 
nevertheless for large but finite times it is impossible to distinguish
signals obtained by (\ref{eq:RW}) and (\ref{eq:mazur}) (see
Fig.~\ref{fig:2}).  This is even more striking 
looking at the computed $\epsilon$-entropy of both signals (see
Fig.~\ref{fig:3}).
\\\indent
The results of Fig.~\ref{fig:3} along with those by Dettman
et al.~\cite{dettman} suggest that, also by assuming the deterministic
character of the system, we are in the practical impossibility of
discerning chaotic from regular motion.
\\\indent
From the above discussion, one may have reached a very pessimistic
view on the possibility to detect the ``true'' nature of a signal by
means of data analysis only. However the scenario is different when
the question about the character of a signal remains restricted only
to a certain interval of scales. In this case, in fact, it is possible
to give an unambiguous classification of the signal character based
solely on the entropy analysis and free from any prior knowledge on
the system/model that generated the data.  Indeed, we can define
stochastic/deterministic behavior of a time series on the basis of the
absence/presence of a saturation plateau $h(\epsilon) \approx
\mbox{const}$ in the observed range of scales.  Moreover the behaviour
of $h(\epsilon,\tau)$ as a function of $(\epsilon,\tau)$ provides a
very useful ``dynamical'' classification of stochastic
processes~\cite{Gaspard93c,PhysicaDnostro}.  One has then a practical
tool to classify the character of a signal as deterministic or
stochastic without referring to a specific model, and is no longer
obliged to answer the metaphysical question, whether the system that
produced the data was a deterministic or a
stochastic~\cite{Kubin95,cenc} one.
\\\indent
This is not a mere way to escape the original question.  Indeed, it is
now clear that the maximum Lyapunov exponent and the Kolmogorov-Sinai
entropy are not completely satisfactory for a proper characterization
of the many faces of complexity and predictability of nontrivial
systems, such as intermittent systems or with many degrees of freedom
(e.g., turbulence)~\cite{rass}.
\\\indent
In the literature the reader may find several methods developed to
distinguish chaos from noise. They are based on the difference in the
predictability using prediction algorithms rather than estimating
the entropy~\cite{Sugihara90a,Casdagli91} or they relate determinism
to the smoothness of the signal~\cite{Kaplan92a,Kaplan93c}. All these
methods have in common the necessity to choose a certain length scale
$\epsilon$ and a particular embedding dimension $m$, therefore 
they suffer the same limitations of the entropy analysis presented here.

\section{Diffusion in non-chaotic systems}
With all the provisos concerning its interpretation, 
Gaspard's and coworkers'~\cite{gaspard} experiment had a very positive 
role not only in stimulating the discussion about the chaos/noise distinction
but also in focusing the attention on deep conceptual aspects of diffusion.
In this context, from a theoretical point of view, the study of 
chaotic models exhibiting diffusion and their non-chaotic 
counterpart contributed to a better understanding of the role of 
chaos on  macroscopic diffusion. 

In Lorentz gases, the diffusion coefficient is related,   
by means of periodic orbits expansion
methods~\cite{gas,dorf,rond}, to chaotic indicators such as the
Lyapunov exponents. This suggested, for certain time, that chaos was or
might have been the basic ingredient for diffusion. However, as argued by 
Dettmann and Cohen~\cite{DC}, even an accurate numerical analysis based on 
the $\epsilon$-entropy has no chance to detect differences in the
diffusive behavior between a chaotic Lorentz gas and its non-chaotic
counterpart, such as the wind-tree Ehrenfest's model.  In the latter model,
particles (wind) scatter against square obstacles (trees) randomly
distributed in the plane but with fixed orientation.  Since the
reflection by the flat edges of the obstacles cannot produce
exponential separation of trajectories, the maximal Lyapunov exponent is
zero and the system is not chaotic. In this case the relation between
the diffusion coefficient and the Lyapunov exponents is of course
nullified.
 
\begin{figure}
\includegraphics[draft=false, scale=0.32, clip=true]{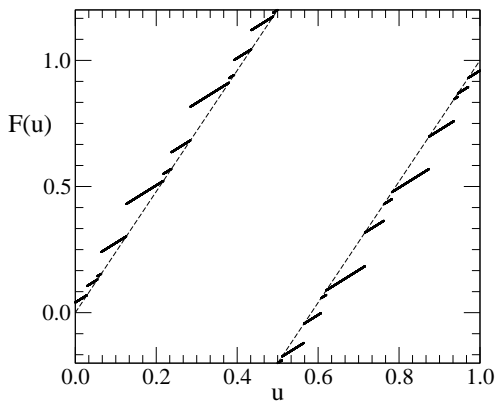}
\caption{Sketch of the random staircase map in the
unitary cell. The parameter $a$ defining the macroscopic slope is set
to $0.23$.  Half domain $[0,1/2]$ is divided into $N=12$
micro-intervals of random size.  The map on $[1/2,1]$ is obtained by
applying the antisymmetric transformation with respect to the center
of the cell $(1/2,1/2)$.}
\label{fig:4}
\end{figure}

The result of Ref.~\onlinecite{DC} implies thus that chaos may be not
indispensable for having deterministic diffusion. The question may be
now posed on what are the necessary microscopic ingredients to observe
deterministic diffusion at large scales.
In the wind-tree Ehrenfest's model, most likely, the disorder in the
distribution of the obstacles is crucial. In
particular, one may conjecture that a finite spatial entropy density,
$h_{S}$, is necessary to the diffusion. So that deterministic
diffusion may be a consequence either of a non-zero ``dynamical''
entropy ($h_{KS}>0$) in chaotic systems or of a non-zero ``static''
entropy ($h_{S} >0$) for non-chaotic systems. This is key-point, 
because someone can argue that a deterministic infinite
system with spatial randomness can be interpreted as an effective
stochastic system, but this is probably a ``matter of taste''.
With the aim of clarifying this point, we consider now a spatially
disordered non-chaotic model~\cite{cecco}, which is the
one-dimensional analog of a two-dimensional non-chaotic Lorentz system
with polygonal obstacles. It has the advantage that both the case with
finite and zero spatial entropy density can be investigated.
Let us start with the map defined by Eqs.~(\ref{eq:chaos}) and
(\ref{eq:chaos2}), and introduce some modifications to make it
non-chaotic.  One can proceed as exemplified by Fig.~\ref{fig:4}, that is 
by replacing the function~(\ref{eq:chaos2}) on each unit cell by its
step-wise approximation that is generated as follows.
The first-half of $C_{\ell}$ is partitioned in $N$ micro-intervals $[\ell
+ \xi_{n-1},\ell + \xi_{n}[$ , $n=1,\dots, N$, with $\xi_0=0 <
\xi_1<\xi_2 < \dots <\xi_{N-1} < \xi_N=1/2$.  In each interval
the map is defined by its linear approximation
\begin{eqnarray}
F_{\Delta}(u)=
\begin{array}{ll}
u - \xi_{n} + F(\xi_n)  \qquad \quad \mbox{if~}
u \in [\xi_{n-1},\xi_{n}[ \;\; ,
\end{array}
\label{model_1}
\end{eqnarray}
where $F(\xi_n) $ is (\ref{eq:chaos2}) evaluated at $\xi_n$.  The map
in the second half of the unit cell is then determined by the
anti-symmetry condition with respect to the middle of the cell.  The
quenched random variables $\{\xi_k\}_{k=1}^{N-1}$ are uniformly
distributed in the interval $[0,1/2]$, i.e. the micro-intervals have a
{\em random} extension. Further they are chosen independently in each
cell $C_{\ell}$ (so one should properly write $\xi_{n}^{(\ell)}$). All
cells are partitioned into the same number $N$ of randomly chosen
micro-intervals (of mean size $\Delta = 1/N$).  The modification of
the continuous chaotic system is conceptually equivalent to replacing
circular by polygonal obstacles in the Lorentz system~\cite{DC}.  The
steps with unitary slope are indeed the analogous of the flat
boundaries of the polygon. While the discontinuities in $F_{\Delta}$,
allowing for a moderate dispersion of trajectories, play a role
similar to the vertex of the polygon that splits a narrow beam of
particles hitting on it.  Since $F_{\Delta}$ has slope $1$ almost
everywhere, the map is no longer chaotic, violating the condition i)
(see Sect.~\ref{sec:diff}).  For $\Delta\to 0$ (i.e. $N\to \infty$)
the continuous chaotic map (\ref{eq:chaos}) is recovered. However,
this limit is singular and as soon as the number of intervals is
finite, even if extremely large, chaos is absent.
It has been found~\cite{cecco}
that this model still exhibits diffusion in the presence of both
quenched disorder and a quasi-periodic external perturbation  
\begin{equation}
x(t+1) = [x(t)] + F_{\Delta}(x(t)-[x(t)]) + \varepsilon \cos(\alpha t)\,.
\label{eq:nochaos}
\end{equation}
The strength of the external forcing is controlled by $\varepsilon$
and $\alpha$ defines its frequency, while $\Delta$ indicates a specific
quenched disorder realization.  

\begin{figure}[thb]
\includegraphics[draft=false, scale=0.32, clip=true]{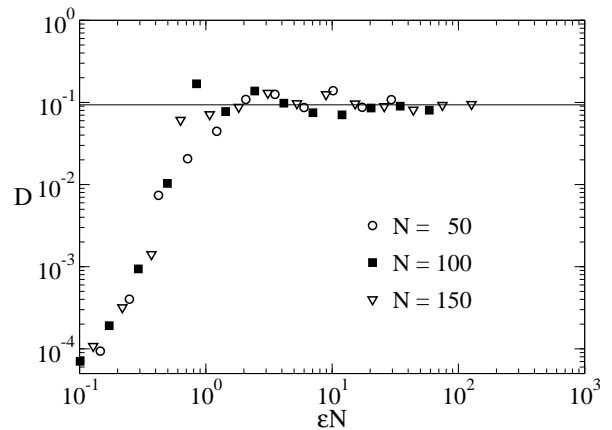}
\caption{Log-Log plot of the dependence of the diffusion coefficient
$D$ on the external forcing strength $\varepsilon$.  Different data
relative to a number of cell micro-intervals $N=50$, $100$ and $150$
are plotted vs the natural scaling variable $\varepsilon N$ to obtain
a collapse of the curves.  Horizontal line represents the result for
chaotic system~(\ref{eq:chaos},\ref{eq:chaos2}).}
\label{fig:5}
\end{figure}

The diffusion coefficient $D$ has been numerically computed from the
linear asymptotic behavior of the mean square displacement. The
results, summarized in Fig.~\ref{fig:5}, show that $D$ is
significantly different from zero only for values
$\varepsilon>\varepsilon_c$.  
For $\varepsilon> \varepsilon_c$, $D$ exhibits a saturation close to
the value of the chaotic system (horizontal line) defined by
Eqs.~(\ref{eq:chaos}) and (\ref{eq:chaos2}).  The existence of a
threshold $\varepsilon_c$ is not surprising.  Due to the staircase
nature of the system, the perturbation has to exceed the typical
discontinuity of $F_{\Delta}$ to activate the ``macroscopic''
instability which is the first step toward the diffusion. Data
collapsing, obtained by plotting $D$ versus $\varepsilon N$, in
Fig.~\ref{fig:5} confirms this argument.  These findings are robust
and do not depend on the details of forcing.  Therefore, we have an
example of a non-chaotic model in the Lyapunov
sense by construction, which performs diffusion.
Now the question concerns the possibility that the diffusive behavior arises 
from the presence of a quenched randomness with non zero spatial entropy 
per unit length.  To clarify this point, similarly to Ref.~\onlinecite{DC}, 
the model can be modified in such a way that the spatial entropy per unit 
cell is forced to be zero, and see if the diffusion still persists.  
 
This programme can be accomplished by repeating the same disorder
configuration every $M$ cells ({\em i.e.}  $\xi^{(\ell)}_n =
\xi^{(\ell+M)}_n$), ensuring a zero entropy for unit length.  Looking
at the diffusion of an ensemble of walkers it was observed that
diffusion is still present with $D$ very close to the expected value
(as in Fig.~\ref{fig:5}).  However, a close analysis reveals the
presence of weak average drift $V$, that vanishes approximately as $V
\sim 1/M$ for large $M$. 
This suggests  that, at large times,  
$\langle (x(t))^2 \rangle \simeq (V t)^2 + 2D t$ 
and the ballistic motion should overcome diffusion. 
However, the crossover time $\tau_c$, at which the motion switches from 
diffusive to ballistic, diverges with $M$ as $\tau_c \sim D M^2$, so 
for very large but finite $M$, the ballistic regime is not observed 
in simulations.
Finally, it should be considered that the value of
$V$ depends on the realization of the randomness, and after averaging
over the disorder the drift becomes zero. Indeed the behavior $V\sim
1/M$ indicates a self-averaging property for large $M$.  
Therefore, we can
conclude that the system displays genuine diffusion for a very long
times even with a vanishing (spatial) entropy density, at least for
sufficiently large $M$.
 
These results along with those by Dettmann and Cohen~\cite{DC} allows
us to draw some conclusions on the fundamental ingredients for observing
deterministic diffusion (both in chaotic and non-chaotic systems).
\begin{itemize}

\item An instability mechanism is necessary to ensure particle
dispersion at small scales (here small means inside the cells). In
chaotic systems this is realized by the sensitivity to the initial
condition.  In non-chaotic systems this may be induced by a finite
size instability mechanisms.  Also with zero maximal Lyapunov exponent
one can have a fast increasing of the distance between two
trajectories initially close~\cite{gpt}. In the wind-tree Ehrenfest
model this stems from the edges of the obstacles, in ``stepwise''
system (\ref{eq:chaos2},\ref{model_1},\ref{eq:nochaos}) from the
jumps.

\item  A mechanisms able to suppress periodic orbits and therefore
to allow for a diffusion at large scale.
\end{itemize}
It is clear that the first requirement is not very strong while the   
the second is more subtle.  In systems with ``strong chaos'', all periodic
orbits are unstable and, so, it is automatically fulfilled.  
In non-chaotic systems, such 
as the non-chaotic billiards studied by Dettmann and Cohen and the map
(\ref{eq:chaos2},\ref{model_1},\ref{eq:nochaos}), the stable periodic
orbits seem to be suppressed or, at least, strongly depressed, by the
quenched randomness (also in the limit of zero spatial entropy).
We note, that, unlike the two dimensional non-chaotic billiards,  
in the 1-d system~(\ref{eq:chaos2},\ref{model_1},\ref{eq:nochaos}),  
the periodic orbits may survive to the presence of disorder, so we need 
the aid of a quasiperiodic perturbation to obtain their destruction and the 
consequent diffusion. 

\section{Discussions and conclusions}
Before summarizing the results of this article, we believe that is
conceptually important to comment about the relevance of chaos in
statistical mechanics approaches.
 
The statistical mechanics~\cite{ehren} had been funded by Maxwell,
Boltzmann and Gibbs for systems with a very large number of degrees of
freedom without any precise requirement on the microscopic
dynamics, apart from the assumption of ergodicity.  After the
discovery of deterministic chaos it becomes clear that also in systems
with few degrees of freedom statistical approaches are necessary.  But,
even after many years, the experts do not agree yet on the fundamental
ingredients which should ensure the validity of the statistical
mechanics.
 
The spectrum of points of view is very wide, ranging from the Landau
(and Khinchin~\cite{khin}) belief on the main role of the many degrees
of freedom and the (almost) complete irrelevance of ergodicity, to the
opinion of who, as Prigogine and his school~\cite{prig1,prig2},
considers chaos as the basic ingredient.  We strongly recommend the
reading of Ref.~\onlinecite{Bricmont} for a detailed discussion of
irreversibility. This work discusses the ``orthodox'' point of view
(based on the role of the large number of degrees of freedom, as
stressed by Boltzmann~\cite{cerci}) re-proposed by
Lebowitz~\cite{lebo} and the following debate on the role of
deterministic chaos~\cite{driebe,prig1}.  Here we focus only on the
aspects related to diffusion problems (and some related aspects,
e.g. conduction).
 
By means of the powerful method of periodic orbits expansion, in
systems with very strong chaos (namely hyperbolic systems), it has
been shown that there exists a close relation between transport
properties (e.g. viscosity, thermal and electrical conductivity and
diffusion coefficients) and indicators of chaos (Lyapunov exponents,
KS entropy, escape rate). These aspects are, e.g., discussed in
Refs.~\onlinecite{gas,dorf}.  At a first glance, the existence of such
relations seem to give evidence against the ``anti dynamical'' point
of view of Landau and Khinchin.
However,
it may be incorrect to employ those results to obtain definite answers
valid for generic systems.
In fact it seems to us that there are rather clear evidences that
chaos is not a necessary condition for the validity of some
statistical behavior~\cite{ford2,R3,Livi}. Beyond the problem of diffusion in
non-chaotic systems, it is worth mentioning the interesting results of Lepri et
al.~\cite{lepri} showing that the Gallavotti-Cohen
formula~\cite{galla}, originally proposed for chaotic systems, holds
also in some non-chaotic model.  Moreover, recently Li and 
coworkers~\cite{li1,li2} studied
the transport properties in quasi-one-dimensional channels with triangular
scatters. In such systems, the maximal Lyapunov exponent is zero
because of the flatness of triangle sides. 
However, numerical simulations
show that, when the scatterers are placed at random (or their height is
random), the Fourier heat law remains valid. Another interesting non-chaotic 
model exhibiting the Fourier heat conduction is the simple 
one-dimensional hard-particle system with alternating masses \cite{GNY}. 
For a recent review on heat conduction in one dimension see 
Ref.\onlinecite{LLP03}).

 These and many other 
examples prove that the heat conduction is present also in system without 
microscopic chaos.  This
is a further indication that microscopic chaos is not the unique
possible source of a macroscopic transport in a given dynamical
system.

Finally let us briefly summarize the main items of this articles.
The problem of distinguishing chaos from noise cannot receive an
absolute answer in the framework of time series analysis.  This is due
to the finiteness of the observational data set and the impossibility
to reach an arbitrary fine resolution and high embedding
dimension. However, this restriction is not necessarily negative, and
we can classify the signal behavior, without referring to any specific
model, as stochastic or deterministic on a certain range of scales.
 
Diffusion may be realized in stochastic and deterministic systems.  In
particular, in the latter case, chaos is not a prerequisite for
observing diffusion and, more in general, nontrivial statistical
behaviors. 

\acknowledgments 
We gratefully thank D. Del-Castillo-Negrete, O.~Kantz and E.~Olbrich
who recently collaborated with us on the issues discussed in this paper. 
We thank G.~Lacorata and L.~Rondoni for very useful remarks on the 
manuscript. 
F.~Cecconi acknowledges the financial support by
FIRB-MIUR RBAU013LSE-001.


\end{document}